\documentclass[
reprint,
 amsmath,amssymb,
 aps,
 prl,
floatfix,
]{revtex4-2}

\usepackage{graphicx}
\usepackage{dcolumn}
\usepackage{bm}
\usepackage{hyperref}
\hypersetup{
    colorlinks=true,
    citecolor=blue,
    linkcolor=blue,  
    urlcolor=blue,
}

\begin{document}

\preprint{APS/123-QED}

\title{Kinetic instability in inductively oscillatory plasma equilibrium}

\author{F. Cruz}
 \email{fabio.cruz@tecnico.ulisboa.pt}
\author{T. Grismayer}
\author{L. O. Silva}
\affiliation{
 GoLP/Instituto de Plasmas e Fus\~{a}o Nuclear, Instituto Superior T\'{e}cnico, Universidade de Lisboa, 1049-001 Lisboa, Portugal
}

\date{\today}

\begin{abstract}
A uniform in space, oscillatory in time plasma equilibrium sustained by a time-dependent current density is analytically and numerically studied resorting to particle-in-cell simulations. The dispersion relation is derived from the Vlasov equation for oscillating equilibrium distribution functions, and used to demonstrate that the plasma has an infinite number of unstable kinetic modes. This instability represents a new kinetic mechanism for the decay of the initial mode of infinite wavelength (or equivalently null wavenumber), for which no classical wave breaking or Landau damping exists. The relativistic generalization of the instability is discussed. In this regime, the growth rate of the fastest growing unstable modes scales with $\gamma_T^{-1/2}$, where $\gamma_T$ is the largest Lorentz factor of the plasma distribution. This result hints that this instability is not as severely suppressed for large Lorentz factor flows as purely streaming instabilities. The relevance of this instability in inductive electric field oscillations driven in pulsar magnetospheres is discussed.
\end{abstract}


\maketitle


Plasma equilibria are usually found as solutions to a combination of kinetic or fluid equations and Maxwell's equations in the stationary limit. Remarkably, time-dependent equilibrium conditions can also be determined for certain systems consisting of a plasma and one or more waves. Such systems are often unstable to parametric instabilities~\cite{nishikawa_1968}, determined by the properties of the waves supporting the equilibrium. Parametric mode excitation can be generally understood as the excitation of two or more plasma waves from a pump wave of finite amplitude $E_0$, frequency $\omega_0$ and wavenumber $k_0$. The pump can be purely electromagnetic, e.g. a laser~\cite{drake_1974, morales_1974, freund_1980, tripathi_1982, dysthe_1983, kaw_dawson_1969, perkins_flick_1971, porkolab_1974}, of Alfv\`{e}nic nature~\cite{bowen_2018, dorfman_2016, matsukiyo_2003, edwards_2016}, or even electrostatic~\cite{zakharov_1972, goldman_1984}. The scope of application of parametric decay is vast, as unstable parametric modes have been explored in the context of laser-plasma interactions~\cite{kaw_dawson_1969, perkins_flick_1971}, inertial~\cite{drake_1974, morales_1974} and magnetic~\cite{porkolab_1974} confinement fusion, laser beam amplification schemes~\cite{forslund_1975} as well as of space and astrophysical plasmas~\cite{weatherall_1981, thejappa_2012, bowen_2018, max_1973}.

In this work, we address the stability of a uniform in space, oscillatory in time plasma equilibrium. In this equilibrium, $(\omega_0, k_0 = 0)$ electric field oscillations, are inductively supported by repeated reversals of the plasma current. This configuration is similar to the oscillating two-stream instability~\footnote{We refer to the oscillating two-stream instability as the instability that develops when two uniform plasma populations (namely electrons and/or positrons) counter-stream with time-dependent average velocities. This is a configuration similar to that in Ref.~\cite{nishikawa_1968}, but in which the pump and all other coupled modes have a similar frequency. This should not be confused with other misleading definitions of the oscillating two-stream instability, discussed in Ref.~\cite{goldman_1984}.}, that has been historically addressed as a parametric instability, but with a time-dependent relative drift velocity~\cite{qin_2014}. This is a regime of interest in pulsar magnetospheres, where inductive oscillations are excited~\cite{levinson_2005, philippov_2020} following electron-positron pair cascades in strong fields~\cite{sturrock_1971, ruderman_sutherland_1975}. In this regime, a fundamental analytical description of the instability is difficult to obtain because different $\omega_0$ harmonics are coupled. Here, we present a theoretical analysis of this instability, and show that it acts as a fundamental plasma process for the transfer of energy from the inductive pump wave to smaller and smaller plasma kinetic scales.

This Letter is organized as follows: first, we describe the plasma equilibrium supporting these oscillations. We then present the dispersion relation of electrostatic plasma waves developed in this equilibrium for an initial waterbag distribution function, and analyse it both theoretically and numerically resorting to particle-in-cell (PIC) simulations. Finally, the conclusions of this work are outlined, and their relevance in pulsar magnetospheres is discussed.

We consider a uniform unmagnetized pair plasma~\footnote{The inductive waves and the instability presented in this work also develop for a single oscillating electron population in an ion background. We choose to present the derivation for pair plasmas due to their applicability to pulsar magnetospheres.} in the presence of a uniform electric field $\mathbf{E} = E \mathbf{\hat{x}} = E_0 \mathbf{\hat{x}}$ which we assume, without loss of generality, to be positive. In the presence of this field, electrons and positrons are accelerated in opposite directions, driving a current $\mathbf{j} = j \mathbf{\hat{x}}$ that inductively reduces $E$. The plasma current is maximum when $E$ vanishes, thus reversing the electric field. Electrons and positrons are decelerated and $j$ decreases in magnitude until it is reversed. The inverse process occurs and the system re-establishes the initial conditions. Since all dynamics is one-dimensional in space, we hereafter restrict our analysis to the $\mathbf{\hat{x}}$ components of fields, currents and particle trajectories. To determine the evolution of the $E$, we first take the time derivative of Amp\`{e}re's law,
\begin{equation}
\frac{\partial^2 E}{\partial t^2} = - 4 \pi e \frac{\partial j}{\partial t}  = - 8 \pi e \int \mathrm{d} p \ v \frac{\partial f_0^+}{\partial t} \ ,
\label{eq:amp_dist}
\end{equation}
where we have assumed that the plasma current is driven by counter-propagating positrons and electrons with uniform density $n_0$ and average velocity $\pm \langle v_+ \rangle$ respectively, \textit{i.e.} $j = 2 e n_0 \langle v_+ \rangle$ ($e$ is the elementary charge). We have also used the definition of average velocity $\langle v_+ \rangle = \int \mathrm{d}p \ v f_0^+ / \int \mathrm{d}p \ f_0^+$, where $f_0^+ = f_0^+(p, t)$ is the positron momentum distribution function, normalized as $\int \mathrm{d}p \ f_0^+(p, t) = n_0$ (the same applies for the electron distribution function $f_0^-$). From the Vlasov equation describing this equilibrium, we can write $\partial f_0^+ / \partial t = - e E \partial f_0^+/\partial p$, and perform the integral in Eq.~\eqref{eq:amp_dist} by parts to obtain
\begin{equation}
\frac{\partial^2 E}{\partial t^2} = - \frac{8 \pi e^2}{m_e} E \int \mathrm{d} p \frac{f_0^+(t, p)}{\gamma^3}  \equiv - \bigg\langle \frac{8\pi e^2 n_0}{m_e \gamma^3} \bigg\rangle E \ ,
\label{eq:amp_eq}
\end{equation}
where we have used the relationship between momentum and velocity, $p = \gamma m_e v$, where $m_e$ is the electron mass and $\gamma = 1/\sqrt{1-v^2/c^2}$. The natural oscillation frequency of the electric field is then $\omega_0 = \sqrt{\langle 8\pi e^2 n_0 / m_e \gamma^3 \rangle} $. In the non relativistic limit, $\omega_0$ is a constant in time and $E$ is purely harmonic. If the electric field amplitude is large enough to accelerate particles to relativistic velocities, the oscillation may be, in general, more complex, with the field changing purely linearly in time between crests and troughs (\textit{i.e.} with a triangular shape)~\cite{levinson_2005}. For simplicity, we present here an analytical description of these waves in non relativistic regime, and then discuss the generalization to the relativistic regime. In the non relativistic limit, the momentum conservation equation of positrons/electrons in the equilibrium defined in Eq.~\eqref{eq:amp_eq} can be integrated, and their unperturbed orbits are
\begin{equation}
v_\pm(t) = v_0 \pm \delta v \cos (\omega_0 t + \theta) \equiv v_0 \pm \delta v \cos \phi \ ,
\label{eq:v_unperturbed}
\end{equation}
where $\delta v = e E_0 / m_e \omega_0$ and $\theta$ is a phase factor such that $v_\pm = v_0$ at a reference time $t = t_0$. We may look for unstable plasma modes $\mathbf{k} = k \mathbf{e}_x$, with $k \neq 0$, that can develop and grow exponentially, by integrating the linearized Vlasov equation along the unperturbed orbits in Eq.~\eqref{eq:v_unperturbed}. This is an approach similar to the derivation of Bernstein waves~\cite{bernstein_1958}, extensively documented in the literature~\cite{stix_1992, keston_2003}. In this Letter, we discuss the final dispersion relation, and present all the details of the derivation in the Supplemental Material.

We consider a plasma where the equilibrium distribution function of both electrons and positrons is a waterbag, $f_0^\pm (v, t) = n_0 / \Delta v ( H(v + v_T \mp \delta v \cos \phi)) - H(v - v_T \mp \delta v \cos \phi) )$, where $\Delta v = 2 v_T$ and $H(x)$ is the Heaviside function. The dispersion relation is
\begin{equation}
1 - \sum_{n=-\infty}^{+\infty} J_n^2 \left( \frac{k \delta v}{\omega_0} \right) \frac{\omega_0^2}{(\omega - n \omega_0)^2 - k^2 v_T^2} = 0 \ ,
\label{eq:disprel_wb}
\end{equation}
where $J_n(x)$ are Bessel functions of the first kind and order $n$. The dispersion relation in Eq.~\eqref{eq:disprel_wb} readily indicates that an infinite number of branches $\omega(k)$ exists. These branches correspond to regions in the $(\omega, k)$ space where each term (in $n$) on the right-hand side of Eq.~\eqref{eq:disprel_wb} dominates the series.

\begin{figure}[t]
\includegraphics[width=3.375in]{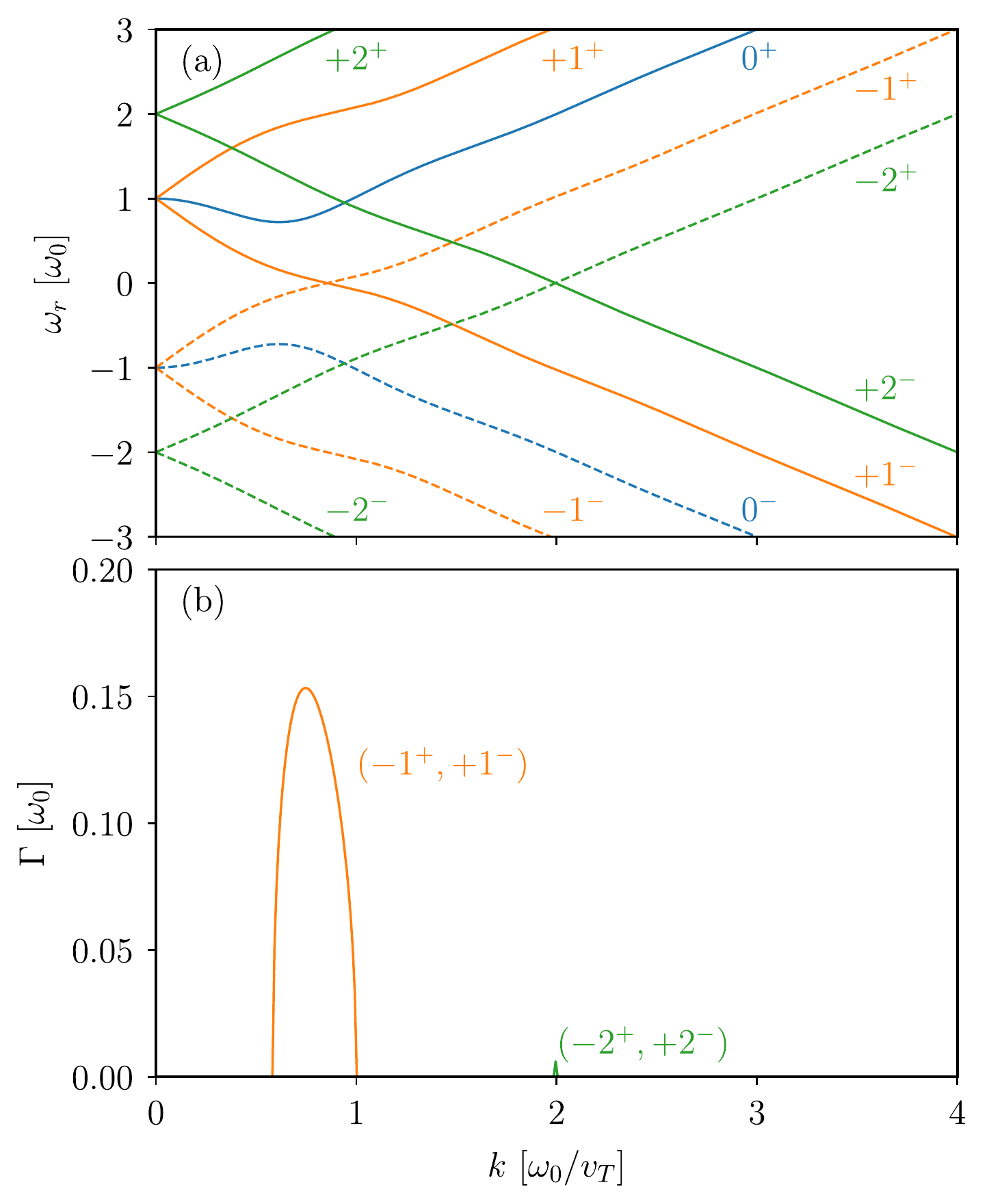}
\caption{\label{fig:disprel_th} Theoretical prediction of real and imaginary components of wave frequencies excited with $\delta v / c = 0.14$ and $\Delta v / c = 0.1$: (a) shows the purely real solutions obtained from individual branches of the full dispersion relation in Eq.~\eqref{eq:disprel_wb}. Lines labeled with $n^\pm$ correspond to solutions where only term $n$ was kept in the series and where $\partial \omega_r / \partial k \gtrless 0$ for $k \gg 1$, respectively; (b) illustrates the growth rate of unstable modes obtained by combining the symmetric branches of Eq.~\eqref{eq:disprel_wb}.}
\end{figure}

Given the complex form of Eq.~\eqref{eq:disprel_wb}, general analytical calculations of the unstable modes and their growth rates are difficult to obtain. Solving Eq.~\eqref{eq:disprel_wb} with each series term individually yields purely real branches, $\omega_{n \pm} = n \omega_0 \pm \sqrt{\omega_0^2 J_n^2 + k^2 v_T^2}$, where $J_n \equiv J_n ( k \delta v / \omega_0)$. The branches with $|n| \leq 2$, relevant for $k v_T / \omega_0 \lesssim 1$ at frequencies near small multiples of $\omega_0$, are plotted in Fig.~\ref{fig:disprel_th}(a) as a function of $k$. An infinite number of crossings between branches exists, with those between consecutive branches $(n, n+1)$ occurring at lower wavenumbers. Symmetric branches $(n, -n)$ cross roughly at $k \simeq n \omega_0 / v_T$ and $\omega_r \equiv \Re (\omega) = 0$ (as expected from symmetry).
Writing $\omega = \omega_r + i \Gamma$ and assuming that $\Gamma \ll \omega_0$, we can show that the terms of the series in Eq.~\eqref{eq:disprel_wb} decrease with $n$, and thus we can keep only the small $n$ terms of the series to solve the dispersion relation. Here, we present two analytical solutions of Eq.~\eqref{eq:disprel_wb} yielding unstable modes, corresponding to the interaction between branches i) $n=0, \pm 1$ at small $k$, and ii) $n=\pm1$ at $k v_T / \omega_0 \simeq 1$. The latter mode only exists for finite $v_T$, whereas the former exists even when $v_T = 0$. For this reason, we hereafter refer to these modes as \textit{thermal} and \textit{fluid}, respectively.

To determine the properties of the thermal mode, we look for solutions to Eq.~\eqref{eq:disprel_wb} with $\omega_r = 0$ with only terms $\pm n$, and then take the particular case $n = 1$. Taking advantage of the Bessel functions' symmetry $J_{-n} = (-1)^n J_n$, we obtain
\begin{equation}
\omega^2_n = n^2 \omega_0^2 + F(k) \pm \sqrt{\omega_0^4 J_n^4 + 4 n^2 \omega_0^2 F(k)} \ ,
\label{eq:disprel_sol}
\end{equation}
where $F(k) = \omega_0^2 J_n^2 + k^2 v_T^2$. The solutions in Eq.~\eqref{eq:disprel_sol} are unstable ($\omega_n^2 < 0$) if $2 \omega_0^2 J_n^2 > n^2 \omega_0^2 - k^2 v_T^2$, which is satisfied for wavenumbers $k v_T / \omega_0 \in (\sqrt{n^2 - J_n^2}, n)$. For $n = 1$ in particular, the symmetric branches cross at $k v_T / \omega_0 = 1$, and $\Gamma_{1, \textrm{max}} \equiv \max (\Im (\omega_1)) \simeq \omega_0 J_1^2 / 2$. If $k \delta v / \omega_0 > 1$, the large argument asymptotic expansion of Bessel functions applies, and we find $\Gamma_{1, \textrm{max}} / \omega_0 \simeq m_e \omega_0 v_T / \pi e E_0 \simeq v_T / \pi \delta v$. In Fig.~\ref{fig:disprel_th}(b), we plot the imaginary component of $\omega_n$ as a function of $k$ for $n = 1, 2$, showing that unstable modes indeed occur close to crossings between the corresponding branches, and that their growth rate decays with $n$.

The fluid mode couples branches $n=0, \pm 1$ at small $k$. For $v_T = 0$ and $k \delta v / \omega_0 \ll 1$, the dispersion relation reduces to
\begin{equation}
\frac{(\Omega - 1)^3}{3\Omega - 1} = K \ ,
\label{eq:disprel_fluid}
\end{equation}
where $\Omega = (\omega / \omega_0)^2$ and $K = (1/2)(k \delta v / \omega_0)^2$. Eq.~\eqref{eq:disprel_fluid} has unstable solutions with $\omega_r  / \omega_0 \simeq \pm 1 \mp (1/4) (k \delta v / \omega_0)^{2/3}$ and growth rate $\Gamma / \omega_0 \simeq (\sqrt{3}/4) (k \delta v / \omega_0)^{2/3}$.

To confirm our theoretical findings, we have performed a set of 1D PIC simulations with OSIRIS~\cite{fonseca_2002, fonseca_2008} considering a uniform pair plasma of density $n_0$ and a waterbag distribution function in momentum with $\Delta v / c = 0.05 - 0.3$. The plasma is subject to an initial electric field $E_0 / (m_e c \omega_p / e) = 0.2$, where $\omega_p^2 = 4 \pi e^2 n_0 / m_e = \omega_0^2 / 2$. The simulation domain has a length $L/(c/\omega_0) \simeq 70$, and is discretized in $N = 5000$ cells, with $500$ particles/cell/species. The simulation time step is $\Delta t \omega_p = 0.005$. When the simulations start, the plasma undergoes the oscillations described by Eq.~\eqref{eq:amp_eq}. Unstable modes grow on top of the $k = 0$ oscillations. These oscillations remain initially stable, but are then damped as energy is transferred to unstable modes and into particle kinetic energy.

\begin{figure*}
\includegraphics[width=7in]{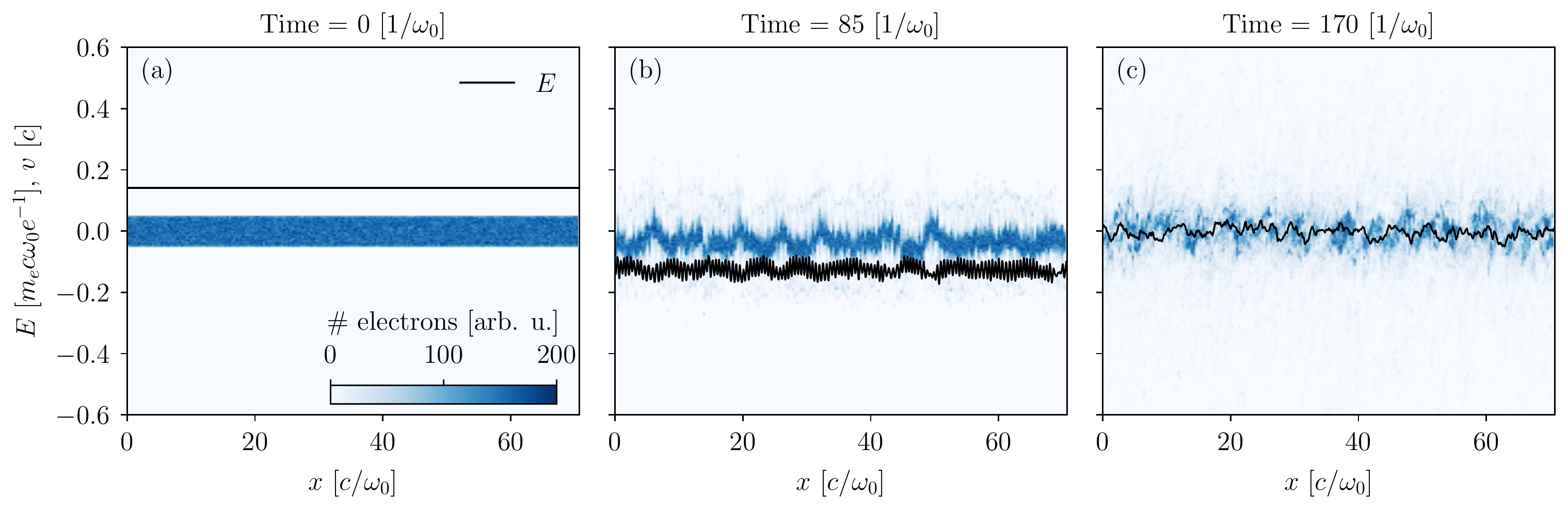}
\caption{\label{fig:phist_e} Temporal evolution of electron phase space (in color) and electric field (in black lines) for a simulation with $E_0/(m_e c \omega_0/ e) = \delta v / c = 0.14$ and $\Delta v / c = 0.1$: (a) initial plasma configuration, (b) linear stage of the instability, showing electric field perturbations grown on top of its uniform oscillatory component, (c) final plasma state, after the instability has saturated.}
\end{figure*}

Fig.~\ref{fig:phist_e} shows the electric field profile and the electron phase space at (a) the beginning of the simulation, (b) during the linear stage of the instability and (c) at a time when the instability has saturated for a simulation with $\Delta v / c = 0.1$. In Fig.~\ref{fig:phist_e}(b), we observe that small perturbations start growing on top of the oscillating electric field, modifying the initial waterbag velocity distribution. The initial electromagnetic energy density ${E_0}^2 / 8 \pi$ is converted into unstable modes until they saturate. The initial particle distribution is then strongly distorted, extending well beyond the initial thermal spread $v_T / c = 0.05$ for long times, see Fig.~\ref{fig:phist_e}(c).

\begin{figure}[t!]
\includegraphics[width=3.375in]{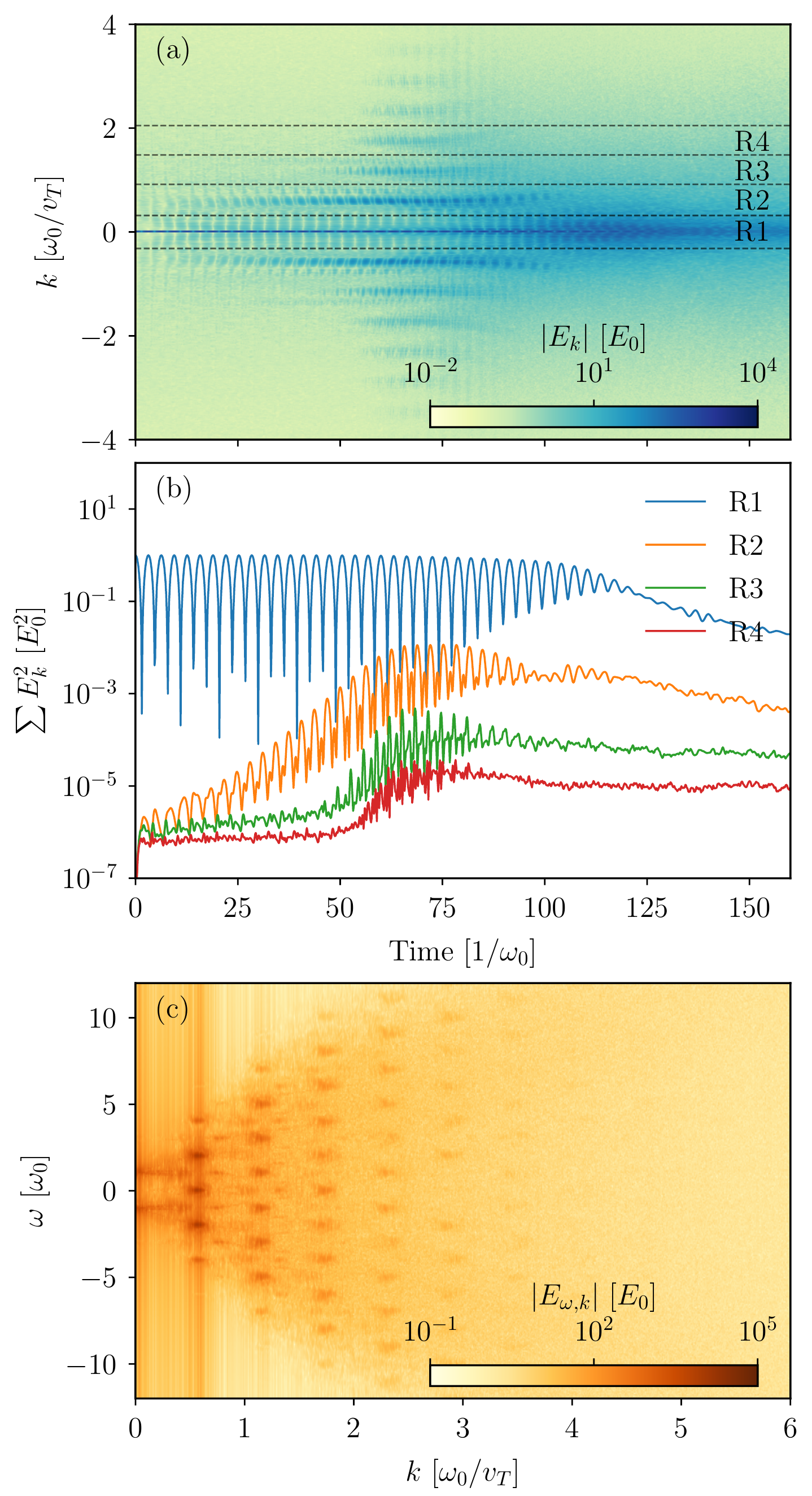}
\caption{\label{fig:ek} Fourier analysis of unstable modes in the simulation illustrated in Fig.~\ref{fig:phist_e}: (a) and (b) show respectively the time evolution of the electric field Fourier transform and of the energy in $k$ bands corresponding to the main oscillatory mode (R1) and to the three unstable modes with lowest $k$ (R2-4); (c) illustrates the numerical plasma dispersion relation, obtained by taking the Fourier transform of (a) in time during the linear stage of the instability.}
\end{figure}

The time evolution of the electric field Fourier spectrum is presented in Fig.~\ref{fig:ek}(a), showing that well-defined unstable modes grow exponentially during the linear stage of the instability. We observe multiple unstable thermal modes: the mode with lowest $k$ is the first to grow (region R2), followed by higher $k$ modes (regions R3-4). We conjecture that these unstable modes are coupled, as discussed in the Supplemental Material. The overall growth rate and total energy stored in unstable modes is dominated by the lowest $k$ thermal mode, as shown in Fig.~\ref{fig:ek}(b), illustrating the electric field energy stored in regions R1-4 of the $k$ space identified in Fig.~\ref{fig:ek}(a). It is also possible to observe in Fig.~\ref{fig:ek}(b) that the energy in R1, corresponding to $k \sim 0$, decreases with time at the expense of the growth of all the other modes. Both the bandwidth and maximum growth rate of the most unstable modes are in good agreement with the solutions of Eq.~\eqref{eq:disprel_sol}. We observe fluid modes grow at $k \delta v / \omega_0 \ll 1$ but only weakly at early times, saturating at levels that do not play any dynamic role on the evolution of the system. When the unstable thermal modes reach a finite amplitude, particle acceleration occurs and causes a saturation of the instability. This is followed by electrostatic turbulence, which causes a strong distortion of the distribution (see Fig.~\ref{fig:phist_e}(c)). The numerical dispersion relation of the plasma in this simulation is presented in Fig.~\ref{fig:ek}(c), which was obtained by Fourier transforming the data in Fig.~\ref{fig:ek}(a) in time during the linear stage of the instability. The numerical dispersion relation shows that the inductive mode is present at $\omega = \omega_0$ for low $k$ and that, in general, unstable modes have a real frequency component multiple of $\omega_0$.

For all simulations performed with $\delta v / v_T \gtrsim 1$, the growth rate of the fastest growing modes is $\Gamma_\textrm{max} / \omega_0 \sim 0.1$ and decreases with increasing $\delta v / v_T$. For $\delta v / v_T \lesssim 1$, we have verified that the growth rate decreases with increasing $v_T$. This is also verified for pair plasmas with Maxwellian distributions with thermal velocities $v_\textrm{th} / c = 0.05 - 0.2$. A detailed discussion of the scaling of the growth rate with  $\delta v / v_T$ is presented in the Supplemental Material. We have found that the fastest growing modes in all simulations with $\delta v / v_T \gtrsim 1$ have $k v_T / \omega_0 \simeq 0.5$, slightly lower than that predicted from linear theory, $k / (\omega_0/v_T) \simeq 0.6 - 1$. We attribute this difference to i) an average performed in the derivation to obtain a dispersion relation independent of time (see Supplemental Material) and to ii) the weakly nonlinear regime in which the thermal modes develop. Regarding ii), we note that fluid modes develop from early times, with which the distribution may interact and broaden slightly, modifying the properties of the thermal mode. For $\delta v / v_T < 1$, simulations show that the fastest growing modes have $k v_T / \omega_0 \sim \delta v / v_T$, a result that the dispersion relation presented in this work fails to explain. Surprisingly, however, the growth rate of these modes is well described by linear theory.

We now discuss the generalization to relativistic conditions, \textit{i.e.} $E_0 / (m_e c \omega_0 / e) \gg 1$. In general, under these conditions, the unperturbed orbits in velocity space become nearly perfect square waves, oscillating between $\pm c$, \textit{i.e.} the trajectories $x(t)$ are triangular waves. It is well known that these can be described as a series of sinusoidal functions that is well approximated by the lowest order term. The same approximation can be applied to the electric field profile in regimes where $\delta p / p_T \gg 1$, where $\delta p = e E_0 / \omega_0$ and $p_T = \gamma_T m_e v_T$ is the momentum thermal spread. In this regime, we find that $\langle 1/\gamma^3 \rangle$ is not constant, and the electric field oscillation is instead well described by a triangular shape~\cite{levinson_2005}. We also performed simulations in the relativistic regime (considering $E_0 / (m_e c \omega_0 / e) \simeq 14$ and $\Delta p / m_e c = 2 - 800$), where all conclusions outlined in this Letter for the non relativistic regime hold qualitatively and quantitatively.

In the relativistic regime, we find also that $\Gamma_\textrm{max}/\omega_0 \sim 0.1$, where $\omega_0 \propto \langle 1 / \gamma^3 \rangle^{1/2}$, according to Eq.~\eqref{eq:amp_eq}. For waterbag and exponential distributions of the type $\exp (- \gamma / \gamma_T)$, we can show that $\langle 1 / \gamma^3 \rangle = \gamma_T^{-1}$ and $(2 \gamma_T)^{-1}$, respectively, and thus $\omega_0 \propto \gamma_T^{-1/2}$. Hence, the unstable modes studied in this work are not as severely suppressed for large $\gamma_T$ as streaming instabilities, for which $\Gamma_\textrm{max} \propto \gamma_T^{-3/2}$, and may be more easily excited in extreme astrophysical settings where the Lorentz factor of the plasma flows is very large. In particular, we find for typical pulsar parameters (surface field $B \simeq 10^{12}$~G, period $0.1$~s) that $\omega_0 \sim 1-10$~GHz~\cite{philippov_2020, cruz_2021}, such that the instability develops on a typical time $1 / \Gamma \sim 10~(p_T / \delta p)^2  / \omega_0 \sim 10~(p_T / \delta p)^2$~ns for the $p_T / \delta p < 1$ regime expected in these scenarios (see Supplemental Material). This should be compared to the lifetime of the inductive plasma waves, which is well approximated by the time between pair production bursts, $T_\textrm{b} \sim 1$~$\mu$s~\cite{timokhin_2010}. We find that $1 / \Gamma < T_\textrm{b}$ for $p_T / \delta p < 10$, which may be achieved in the plasma trail behind pair production fronts, where the electric field oscillates inductively, but at an amplitude that is not enough to trigger a considerable number of pair production events~\cite{cruz_2021}. This instability may also perturb particle trajectories in inductive waves, previously identified as a source of linear acceleration emission~\cite{melrose_2009b, reville_2010}. We note that the strong longitudinal magnetic field typical of these environments is not expected to play a significant role in the development of this instability (contrary to e.g. filamentation modes), as particle trajectories are purely one-dimensional.

In conclusion, we have studied the plasma waves parametrically excited in an inductively oscillatory plasma equilibrium. Our results show that the energy in the inductive pump wave is transferred to other plasma modes via an oscillating two-stream instability. Since the pump wave has infinite wavelength, other fundamental wave depletion mechanisms such as wave breaking or Landau damping do not operate, regardless of the value of its amplitude $E_0$. We have presented the dispersion relation of these waves for a waterbag equilibrium distribution function, which captures thermal effects and is analytically tractable. In general, infinite branches $\omega(k)$ exist, each branch being a purely real mode when far from other branches in the $(\omega, k)$ space. However, coupling between different branches yields unstable modes, with the maximum growth rate $\Gamma_\textrm{max} / \omega_0 \sim 0.1$. All analytical results have been confirmed using PIC simulations. We have also investigated the relevance of this instability in pulsar magnetospheres, and determined that it may be excited over short distances following pair production bursts in strong fields. Furthermore, we speculate that it may be relevant in other astrophysical scenarios, e.g. black hole magnetospheres~\cite{levinson_2018, chen_2020b, kisaka_2020}, where strong rotation-powered electric fields are also self-consistently screened by plasma currents.

\begin{acknowledgments}
F. Cruz and T. Grismayer contributed equally to this work. This work was supported by the European Research Council (ERC-2015-AdG Grant 695088) and FCT (Portugal) (grant PD/BD/114307/2016) in the framework of the Advanced Program in Plasma Science and Engineering (APPLAuSE, FCT grant PD/00505/2012). We acknowledge PRACE for granting access to MareNostrum, Barcelona Supercomputing Center (Spain), where the simulations presented in this work were performed.
\end{acknowledgments}

\newpage
\ 
\newpage
\
\newpage

\onecolumngrid
\begin{center}
\Large{\textbf{Supplemental Material}}
\end{center}

\renewcommand{\thefigure}{S\arabic{figure}}
\renewcommand{\theequation}{S\arabic{equation}}

\setcounter{equation}{0}
\setcounter{figure}{0}

\section{\label{app:derivation}Derivation of the dispersion relation}

We start by linearizing the Vlasov equation for positrons and electrons (superscripts $\pm$, respectively),
\begin{equation}
\frac{\partial f_1^\pm}{\partial t} + v_\pm \frac{\partial f_1^\pm}{\partial x} \pm \frac{e E_0}{m_e} \frac{\partial f_1^\pm}{\partial v} = \mp \frac{e E_1}{m_e} \frac{\partial f_0^\pm}{\partial v} \ ,
\label{eq:lin_vlasov}
\end{equation}
where subscripts $0$ ($1$) correspond to zeroth (first) order, time-varying quantities. The left hand side of Eq.~\eqref{eq:lin_vlasov} can be interpreted as a total time derivative of $f_1^\pm$ in phase space. Thus, $f_1^\pm$ can be obtained by integrating the right hand side of Eq.~\eqref{eq:lin_vlasov},
\begin{equation}
f_1^\pm (x, v, t) = \mp \frac{e}{m_e} \int_{-\infty}^t \mathrm{d} t' E_1(x', t') \frac{\partial f_0^\pm (x', v', t')}{\partial v'} \ ,
\label{eq:lin_vlasov_sol}
\end{equation}
where all primed quantities are taken along the unperturbed trajectories. The solution in Eq.~\eqref{eq:lin_vlasov_sol} can then be used in Poisson's equation to find the dispersion relation. Using an anzats of the type $A_1 \sim \bar{A}_1 \exp (i (k x - \omega t))$ for first order quantities, we can write
\begin{equation}
i k \bar{E}_1 = -\sum_{s=\pm} \frac{4 \pi e^2}{m_e} \int \mathrm{d} v \int_{-\infty}^t \mathrm{d} t' \bar{E}_1 \frac{\partial f_0^s}{\partial v'} \exp \left[ i k (x'_s-x_s) - i \omega (t' - t)] \right] \ .
\label{eq:disprel_ini}
\end{equation}
We now attempt to calculate the integrals in Eq.~\eqref{eq:disprel_ini} along the unperturbed trajectories. These can be written, in the non relativistic regime, as
\begin{equation}
v_\pm(t) = v_0 \pm \delta v \cos (\omega_0 t + \theta) \equiv v_0 \pm \delta v \cos \phi \ ,
\label{eq:v_unperturbed_app}
\end{equation}
where the phase $\theta$ is such that $v = v_0$ at some reference time $t_0$, and is the same for all particles in the distribution. Eq.~\eqref{eq:v_unperturbed_app} allows us to write the velocity at any time $t'$ as
\begin{equation}
v'_\pm \equiv v_\pm(t') = v_0 \pm \delta v \cos(\omega_0(t'-t) + \phi) = v_\pm \pm \delta v [ \cos(\omega_0(t'-t) + \phi) - \cos \phi ]  \ ,
\label{eq:vprime_unperturbed}
\end{equation}
and the difference in positions between times $t'$ and $t$ as
\begin{equation}
x'_\pm - x_\pm = \int_t^{t'} \mathrm{d} \tau v(\tau) = v_0 (t'-t) \pm \frac{\delta v}{\omega_0} [ \sin(\omega_0(t'-t) + \phi) - \sin \phi ]  \ .
\label{eq:xprime_unperturbed}
\end{equation}
Using the difference in Eq.~\eqref{eq:xprime_unperturbed}, we can write the exponential term in Eq.~\eqref{eq:disprel_ini} for positrons as
\begin{equation}
\exp \left[ - i (\omega - k v_0) (t' - t) \right] \exp \left[ i k \delta v / \omega_0 \left( \sin(\omega_0 (t'-t) + \phi) - \sin \phi \right) \right] \ .
\label{eq:exp_factors}
\end{equation}
For the electron species, this exponential term has the same shape, with $\delta v \to - \delta v$. This difference does not introduce any change in the final result, so we proceed with the derivation focusing on the positron species terms. Let us focus on the last exponential in Eq.~\eqref{eq:exp_factors}. Using the Bessel identity
\begin{equation*}
e^{i a \sin x} = \sum_{n=-\infty}^{+\infty} J_n(a) e^{inx} \ ,
\end{equation*}
where $J_n$ is the Bessel function of the first kind and order $n$, we can write the last exponential in Eq.~\eqref{eq:exp_factors} as
\begin{equation}
\sum_{n=-\infty}^{+\infty} \sum_{m=-\infty}^{+\infty} J_n \left( \frac{k\delta v}{\omega_0} \right) J_m \left( \frac{k\delta v}{\omega_0} \right) \exp[ i n \omega_0 (t' - t)] \exp[ i (n-m) \phi ] \ .
\end{equation}
This can be plugged back in Eq.~\eqref{eq:disprel_ini} to obtain
\begin{equation}
i k = -\sum_s \frac{4 \pi e^2}{m_e} \int \mathrm{d} v \int_{-\infty}^t \mathrm{d} t' \frac{\partial f_0^s}{\partial v'} \sum_{n,m} J_n J_m \exp \left[ - i (\omega - n \omega_0 - k v_0)(t'-t) \right] A_{n,m} (\phi)  \ ,
\label{eq:disprel_sums}
\end{equation}
where we have written $A_{n,m}(\phi) = \exp \left[ i (n-m) \phi \right]$ and where we have omitted the limits of the sums in $n$ and $m$ and the arguments of the Bessel functions $J_n$ and $J_m$. We now focus on evaluating the integrals in Eq.~\eqref{eq:disprel_sums}, by considering, for simplicity, the zeroth order distribution functions as waterbags,
\begin{equation}
f_0^\pm \equiv f_0^\pm(v, t) = \frac{n_0}{\Delta v} \left[ H(v + v_T \mp \delta v \cos \phi) - H(v - v_T \mp \delta v \cos \phi) \right] \ .
\label{eq:wb_dist}
\end{equation}
Here, $\Delta v = 2 v_T$ is the thermal spread of the distribution and $H(x)$ is the Heaviside function. The derivative with respect to $v$ of the distribution function reads $\partial f_0^\pm / \partial v = n_0 / \Delta v (\delta(v + v_T \mp \delta v \cos \phi) - \delta(v - v_T \mp \delta v \cos \phi))$, where $\delta(x) = \mathrm{d}H(x)/\mathrm{d}x$.

To evaluate the integral in $v$ in Eq.~\eqref{eq:disprel_sums}, we note from Eq.~\eqref{eq:v_unperturbed_app} and \eqref{eq:wb_dist}, that the Dirac deltas in $\partial f_0 / \partial v'$ are of the type
\begin{equation}
\delta ( v' + v_T \mp \delta v \cos(\omega_0(t'-t) + \phi) ) = \delta ( v + v_T \mp \delta v \cos \phi ) \ ,
\end{equation}
and thus $\partial f_0 / \partial v' = \partial f_0 / \partial v$. We can finally replace $v_0 = v \mp \delta v \cos \phi$ for positrons and electrons, respectively, and the integration in $v$ can be easily performed. The dispersion relation becomes, at this point,
\begin{align}
 i k = & - \frac{8 \pi e^2 n_0}{m_e \Delta v} \int_{-\infty}^t \mathrm{d} t' \sum_{n,m} J_n J_m A_{n,m} (\phi) \times \nonumber \\
 & \big\{ \exp \left[ - i (\omega - n \omega_0 + k v_T)(t'-t) \right] - \exp \left[ - i (\omega - n \omega_0 - k v_T)(t'-t) \right] \big\} \ ,
\label{eq:disprel_exps}
\end{align}
where we have performed the sum over species. We now change the time integration variable to $\tau = t - t'$. In this case, we have
\begin{align}
  i k =  & \frac{8 \pi e^2 n_0}{m_e \Delta v} \int_0^{+\infty} \mathrm{d} \tau \sum_{n,m} J_n J_m A_{n,m} (\phi) \times \nonumber \\
 & \big\{ \exp \left[ i (\omega - n \omega_0 + k v_T) \tau \right] - \exp \left[ i (\omega - n \omega_0 - k v_T) \tau \right] \big\} \ .
\label{eq:disprel_exps_tau}
\end{align}
The integrals in Eq.~\eqref{eq:disprel_exps_tau} can be readily evaluated, and we find
\begin{equation}
1 - \sum_n J_n \frac{\omega_0^2}{(\omega - n \omega_0)^2 - k^2 v_T^2} \sum_m J_m A_{n,m} (\phi) = 0 \ .
\label{eq:disprel_exps_anm}
\end{equation}
To obtain a dispersion relation that does not depend on $\phi$, we take an average over one period of the driving oscillation of $A_{n, m}$,
\begin{equation}
\langle A_{n,m}(\phi) \rangle = \frac{1}{2\pi} \int_0^{2\pi} \mathrm{d} \phi A_{n,m}(\phi) = \frac{1}{2\pi} \int_0^{2\pi} \mathrm{d} \phi \exp[ i (n-m) \phi ] = \delta_{n,m} \ ,
\end{equation}
where we have used the definition of the Kronecker delta $\delta_{n,m}$. Plugging this in Eq.~\eqref{eq:disprel_exps_anm}, the dispersion relation reduces to
\begin{equation}
1 - \sum_{n=-\infty}^{+\infty} J_n^2 \left( \frac{k \delta v}{\omega_0} \right) \frac{\omega_0^2}{(\omega - n \omega_0)^2 - k^2 v_T^2} = 0 \ .
\end{equation}

\section{\label{app:dist}Mode coupling equations}


The dispersion relation obtained analytically and numerically for the instability described in this manuscript hints that unstable modes are coupled. Here we derive a closed set of equations in the fluid limit that puts in evidence this coupling, and use it to interpret why some modes may dominate in the simulations presented in this work. For simplicity, we present this derivation in the non relativistic regime. We attribute the subscripts $0$, $1$ to zeroth order (equilibrium) and first order quantities and the superscripts $\pm$ to quantities associated with positrons and electrons, respectively. We start by linearizing the continuity equations,
\begin{equation}
\frac{\partial n_1^\pm}{\partial t} \pm v_0(t) \frac{\partial n_1^\pm}{\partial x} + n_0 \frac{\partial v_1^\pm}{\partial x} = 0 \ . 
\label{eq:continuity}
\end{equation}
Here, $v_0(t)$ is the time dependant zeroth order fluid velocity developed by both species in the oscillating zeroth order electric field $E_0(t)$. Taking the partial time derivative of Eq.~\eqref{eq:continuity}, we get
\begin{equation}
\frac{\partial^2 n_1^\pm}{\partial t^2} \pm \frac{\partial v_0}{\partial t} \frac{\partial n_1^\pm}{\partial x} \pm v_0 \frac{\partial^2 n_1^\pm}{\partial t \partial x} + n_0 \frac{\partial^2 v_1^\pm}{\partial t \partial x} = 0 \ . 
\end{equation}
Changing the order of the derivatives in the last term, and using the momentum equation, we have
\begin{equation}
\frac{\partial^2 n_1^\pm}{\partial t^2} \pm \frac{\partial v_0}{\partial t} \frac{\partial n_1^\pm}{\partial x} \pm v_0 \frac{\partial^2 n_1^\pm}{\partial t \partial x} + n_0 \frac{\partial}{\partial x} \left( - \frac{1}{n_0 m_e} \frac{\partial p^\pm}{\partial x} \pm \frac{e}{m_e} E_1 \mp v_0 \frac{\partial v_1^\pm}{\partial x} \right) = 0 \ ,
\label{eq:continuity_terms}
\end{equation}
where $p^\pm$ is the positron/electron fluid pressure. The last two terms of Eq.~\eqref{eq:continuity_terms} can be expressed as a function of $n_1^\pm$ by using Gauss's law and the continuity equations, respectively,
\begin{equation}
\frac{\partial^2 n_1^\pm}{\partial t^2} \pm \frac{\partial v_0}{\partial t} \frac{\partial n_1^\pm}{\partial x} \pm v_0 \frac{\partial^2 n_1^\pm}{\partial t \partial x} \pm \omega_p^2 (n_1^+ - n_1^-) - \frac{\gamma^\pm T^\pm}{m_e} \frac{\partial^2 n_1^\pm}{\partial x^2} \mp v_0 \frac{\partial}{\partial x} \left(  -\frac{\partial n_1^\pm}{\partial t} \mp v_0 \frac{\partial n_1^\pm}{\partial x} \right) = 0 \ ,
\label{eq:continuity_cmplx}
\end{equation}
where we have written $\omega_p^2 = 4 \pi e^2 n_0 / m_e$ and $p^\pm = \gamma^\pm n^\pm T^\pm$, with $\gamma^\pm$ and $T^\pm$ being the adiabatic index and temperature of each fluid, respectively. Eq.~\eqref{eq:continuity_cmplx} can be simplified to
\begin{equation}
\frac{\partial^2 n_1^\pm}{\partial t^2} + \left( v_0^2 - \gamma^\pm {v_\textrm{th}^\pm}^2 \right) \frac{\partial^2 n_1^\pm}{\partial x^2} \pm \frac{\partial v_0}{\partial t} \frac{\partial n_1^\pm}{\partial x} \pm 2 v_0 \frac{\partial^2 n_1^\pm}{\partial t \partial x} \pm \omega_p^2 (n_1^+ - n_1^-) = 0 \ , 
\label{eq:continuity_simp}
\end{equation}
with ${v_\textrm{th}^\pm}^2 = T^\pm / m_e$. Eq.~\eqref{eq:continuity_simp} describes two non-trivially forced and coupled oscillators, $n_1^\pm$. Assuming now that $v_0(t) = \delta v \cos (\omega_0 t) = \delta v / 2 (\exp(i \omega_0 t) + \exp(-i \omega_0 t))$, we can look for wave solutions to Eq.~\eqref{eq:continuity_simp}, which is a generalization of Mathieu's equation. For a thorough review of the stability properties of Mathieu's equation, see ref.~\cite{kovacic_2018}.

A particularly insightful equation can be obtained by taking the Fourier transform in space and time of Eq.~\eqref{eq:continuity_simp}. After some algebra, we obtain
\begin{equation}
A_0^\pm n_1^\pm(\omega) \pm A_{1+} n^\pm_1 (\omega+\omega_0) \pm A_{1-} n^\pm_1 (\omega - \omega_0) - A_2 \left[ n^\pm_1(\omega+2\omega_0) + n^\pm_1(\omega-2\omega_0) \right] = \omega_p^2 n^\mp_1 (\omega) \ ,
\label{eq:coupling}
\end{equation}
where $n_1^\pm(\omega) = n_1^\pm(\omega, k)$ is the Fourier mode with frequency $\omega$ and wave vector $k$, and where $A_0^\pm = \omega_p^2 + \gamma^\pm k^2 {v_\textrm{th}^\pm}^2 - k^2 \delta v^2 / 2 - \omega^2$, $A_{1\pm} = (k \delta v / 2) (2\omega \pm \omega_0)$ and $A_2 = (k \delta v / 2)^2$. Eq.~\eqref{eq:coupling} shows that each mode $\omega$ is coupled to its neighbours $\omega \pm \omega_0$ and $\omega \pm 2\omega_0$. This result is similar to that obtained in other works describing multiple light/plasma wave interactions originally inspired by ref.~\cite{nishikawa_1968}.

A possible approach to obtain a dispersion relation from Eq.~\eqref{eq:coupling} is to solve the system of coupled equations for neighbour modes, e.g. $\omega$, $\omega \pm \omega_0$ and $\omega \pm 2 \omega_0$. This yields a system of linear equations that can be truncated to any desired neighbour mode order. This truncation corresponds to an ordering condition in $k \delta v / \omega_0$, since $A_{1 \pm} \propto k \delta v / \omega_0$ and $A_1 \propto (k \delta v / \omega_0)^2$. We have verified numerically that keeping only modes $\omega$ and $\omega \pm 1$ yields a system of linear equations whose solution recovers the dispersion relation in Eq.~\eqref{eq:disprel_fluid}, and in particular the scaling $\Gamma / \omega_0 \propto (k \delta v / \omega_0)^{2/3}$.


\section{\label{app:scaling}Scaling of instability growth rate with temperature}

\begin{figure}[t]
\includegraphics[width=3.375in]{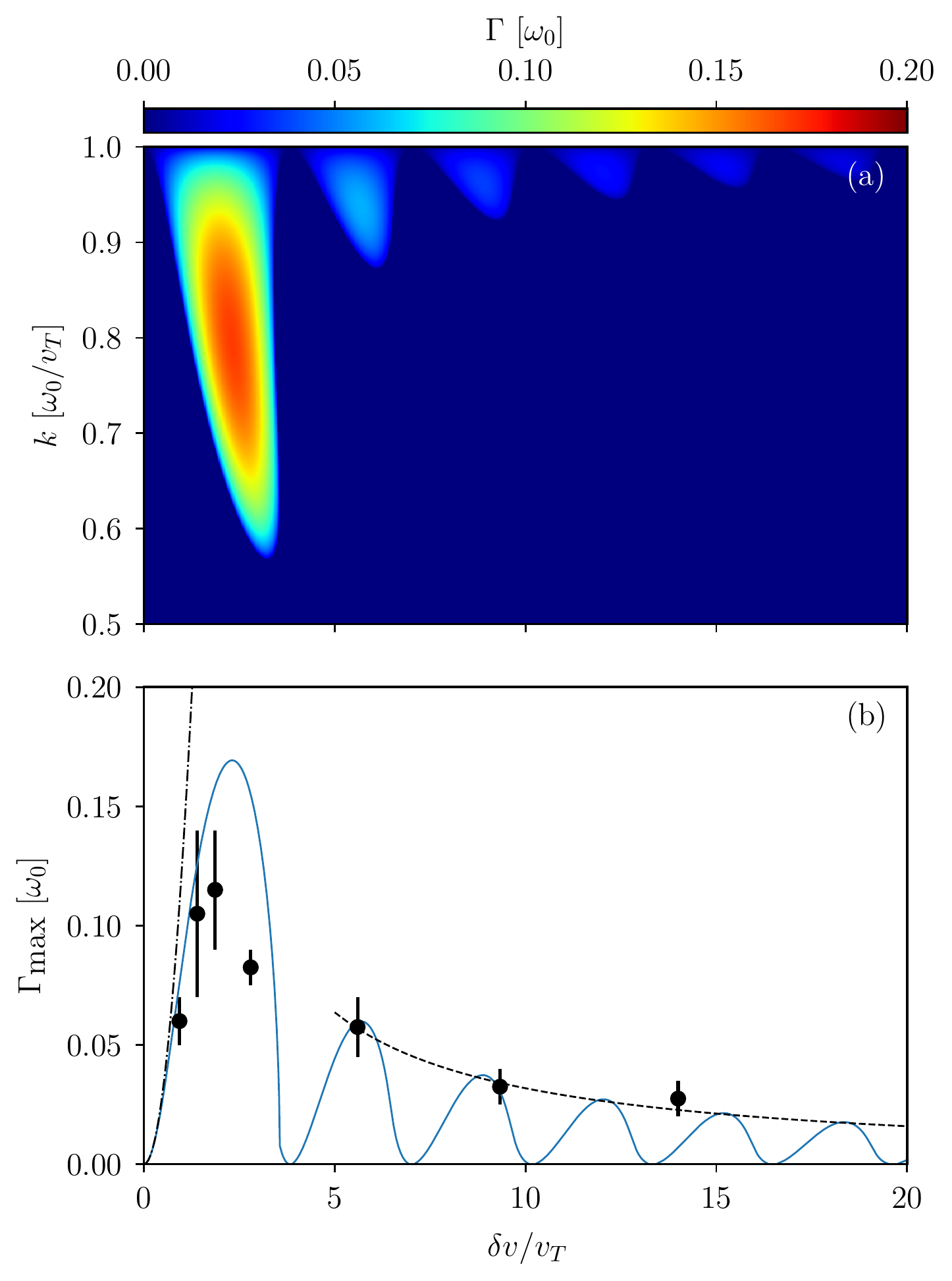}
\caption{\label{fig:grate} Instability growth rate. (a) shows the wavenumber and $\delta v / v_T$ ratio dependence of the growth rate, and (b) illustrates the dependence of the maximum growth rate for fixed $\delta v / v_T$ as a function of the wavenumber. Dashed-dotted and dashed black lines in (b) represent theoretical asymptotic limits of the maximum growth rate, whereas black dots represent simulation results.}
\end{figure}

We discuss here the scaling with temperature of the growth rate of the instability presented in this manuscript. For simplicity, we adopt the normalization $\delta v \to \delta v / c$, $v_T \to v_T/c$, $k \to k v_T / \omega_0$ and $\omega \to \omega / \omega_0$. The dispersion relation in Eq.~\eqref{eq:disprel_wb} reads
\begin{equation}
1 - \sum_n J_n^2 \left( k \frac{\delta v}{ v_T} \right) \frac{1}{(\omega - n)^2 - k^2} = 0 \ .
\label{eq:disprel_wb_norm}
\end{equation}
As mentioned in the manuscript, keeping only the terms $n = \pm 1$ in this series is a reasonable approximation to determine the growth rate of unstable thermal modes. The solution to Eq.~\eqref{eq:disprel_wb_norm} is, in this approximation, given by
\begin{equation}
\omega^2 = 1 + F(k) \pm \sqrt{ J_1^4(k\delta v/v_T) + 4 F(k)} \ ,
\label{eq:disprel_sol_norm}
\end{equation}
with $F(k) = J_n^2(k\delta v/v_T) + k^2$. The imaginary component of this solution, $\Im (\omega) \equiv \Gamma$, is plotted in Figure~\ref{fig:grate}(a) as a function of $k$ and $\delta v / v_T$, and its maximum value for a given $\delta v / v_T$ ratio, $\Gamma_\textrm{max}$, is plotted in Figure~\ref{fig:grate}(b) as a function of $k$. In Figure~\ref{fig:grate}(b) we show also in dashed-dotted and dashed black lines the analytical estimate in the asymptotic limits of small and large $\delta v / v_T$. These lines are obtained by taking asymptotic limits of the Bessel function $J_1$, with which we can estimate $\Gamma_\textrm{max} \simeq \omega_0 J_1^2 / 2$. For $\delta v / v_T \gg 1$, we have $\Gamma_\textrm{max} / \omega_0 \simeq v_T / \pi \delta v$, whereas for $\delta v / v_T \ll 1$ we find $\Gamma_\textrm{max} / \omega_0 \simeq (\delta v / v_T)^2 / 8$. Figure~\ref{fig:grate}(b) also shows the growth rate of the fastest growing modes in several simulations with different $\delta v / v_T$ ratios in black dots, showing a good agreement with theoretical predictions. All simulations were performed with fixed $\delta v / c \simeq 0.14$ and varying $v_T$. The growth rates were obtained by plotting the time evolution of the energy in a region in $k$ space in the vicinity of the fastest growing modes (see Figure~\ref{fig:ek}(a) and (b)), and fitting exponential functions in the linear phase of the instability.

\nocite{*}
\bibliography{apssamp} 
\end{document}